\documentclass[12pt]{article}
\usepackage{graphicx}
\usepackage{amsmath}
\def\Re{{\cal R \mskip-4mu \lower.1ex \hbox{\it e}\,}}
\def\Im{{\cal I \mskip-5mu \lower.1ex \hbox{\it m}\,}}
\def\ie{{\it i.e.}}
\def\eg{{\it e.g.}}
\def\etc{{\it etc}}

\def\sub#1{_{\lower.25ex\hbox{$\scriptstyle#1$}}}

\def\to{\rightarrow}

\def\subw{_{\rm w}}
\def\mh{\ifmmode m\sbl H \else $m\sbl H$\fi}
\def\mch{\ifmmode m_{H^\pm} \else $m_{H^\pm}$\fi}
\def\mt{\ifmmode m_t\else $m_t$\fi}
\def\mc{\ifmmode m_c\else $m_c$\fi}
\def\mz{\ifmmode M_Z\else $M_Z$\fi}
\def\mw{\ifmmode M_W\else $M_W$\fi}
\def\mws{\ifmmode M_W^2 \else $M_W^2$\fi}
\def\mhs{\ifmmode m_H^2 \else $m_H^2$\fi}   
\def\mzs{\ifmmode M_Z^2 \else $M_Z^2$\fi}
\def\mts{\ifmmode m_t^2 \else $m_t^2$\fi}
\def\mcs{\ifmmode m_c^2 \else $m_c^2$\fi}
\def\mchs{\ifmmode m_{H^\pm}^2 \else $m_{H^\pm}^2$\fi}
\def\ztwo{\ifmmode Z_2\else $Z_2$\fi}
\def\zone{\ifmmode Z_1\else $Z_1$\fi}
\def\mtwo{\ifmmode M_2\else $M_2$\fi}
\def\mone{\ifmmode M_1\else $M_1$\fi}
\def\tb{\ifmmode \tan\beta \else $\tan\beta$\fi}
\def\xw{\ifmmode x\subw\else $x\subw$\fi}
\def\ch{\ifmmode H^\pm \else $H^\pm$\fi}
\def\lum{\ifmmode {\cal L}\else ${\cal L}$\fi}
\def\inpb{\ifmmode {\rm pb}^{-1}\else ${\rm pb}^{-1}$\fi}
\def\infb{\ifmmode {\rm fb}^{-1}\else ${\rm fb}^{-1}$\fi}
\def\epem{\ifmmode e^+e^-\else $e^+e^-$\fi}
\def\ppb{\ifmmode \bar pp\else $\bar pp$\fi}
\def\pbp{\ifmmode ~^(\bar p^)p\else $~^(\bar p^)p$\fi}
\def\bsg{\ifmmode B\to X_s\gamma\else $B\to X_s\gamma$\fi}
\def\bsll{\ifmmode B\to X_s\ell^+\ell^-\else $B\to X_s\ell^+\ell^-$\fi}
\def\bstt{\ifmmode B\to X_s\tau^+\tau^-\else $B\to X_s\tau^+\tau^-$\fi}

\newskip\zatskip \zatskip=0pt plus0pt minus0pt
\def\matth{\mathsurround=0pt}

\def\atversim#1#2{\lower0.7ex\vbox{\baselineskip\zatskip\lineskip\zatskip
  \lineskiplimit 0pt\ialign{$\matth#1\hfil##\hfil$\crcr#2\crcr\sim\crcr}}}

\renewcommand{\thefootnote}{\fnsymbol{footnote}}

\begin{document} \begin{titlepage} 
\rightline{\vbox{\halign{&#\hfil\cr
&SLAC-PUB-14768\cr
}}}
\vspace{1in} 
\begin{center}

{{\Large\bf Gauge Kinetic Mixing in the $E_6$SSM}
\footnote{Work supported by the Department of 
Energy, Contract DE-AC02-76SF00515}\\}
\medskip
\medskip
\normalsize 
{\large Thomas G. Rizzo{\footnote {e-mail: rizzo@slac.stanford.edu}} \\
\vskip .6cm
SLAC National Accelerator Laboratory,  \\
2575 Sand Hill Rd, Menlo Park, CA 94025, USA\\}
\vskip .5cm

\end{center} 
\vskip 0.8cm

\begin{abstract} 

The $E_6$SSM extension of the MSSM allows for the solution of many of the difficulties that are usually encountered within conventional SUSY 
breaking scenarios, \eg, the $\mu$ problem, the imposition of R-parity `by hand', the generation of light neutrino masses and obtaining a light 
Higgs boson with a mass as large as $\sim$ 125 GeV as suggested by recent LHC measurements. In addressing these problems, such a scenario predicts 
the existence of additional singlet and vector-like superfields beyond those in the MSSM as well as possibly two new neutral gauge bosons near 
the TeV scale. In this paper the phenomenological implications of simultaneous gauge kinetic mixing between the usual Standard Model (SM) 
hypercharge gauge field and both these new neutral gauge fields present in the $E_6$SSM scenario is explored. To this end a large class of specific 
toy models realizing this type of kinetic mixing is examined. In particular, we demonstrate that a significant suppression (or enhancement) of the 
expected event rate for $Z'$ production in the dilepton channel at the LHC is not likely to occur in this scenario due to gauge kinetic mixing.

\end{abstract}

\renewcommand{\thefootnote}{\arabic{footnote}} \end{titlepage}

%%%%%%%%%%%%%%%%%%%%%%%%%%%%%%%---- Put text here

\section{Introduction and Background}

Supersymmetry, which is softly broken at the TeV scale, is one of the most attractive extensions of the Standard Model (SM) as it directly addresses 
the hierarchy problem, provides interesting dark matter candidates and allows for grand unification of the SM gauge couplings. However, 
the simplest version of SUSY, the MSSM{\cite {Drees:2004jm}}, has a number of outstanding non-trivial difficulties associated with it which include: 
the need to impose an R-parity-like 
symmetry to avoid rapid proton decay, the lack of a mechanism to generate light neutrino masses, and the so-called $\mu$-problem, \ie, how to 
naturally obtain values of the $\mu$ parameter at or below the $\sim 1$ TeV mass scale. Furthermore, in some of the well known SUSY-breaking 
scenarios, it is somewhat difficult{\cite {implications}} within the MSSM to generate a sufficiently large light Higgs mass in the neighborhood of 
$\simeq 125$ GeV, as is suggested by the recent results from the LHC{\cite {lhcnotes}}, without significant fine-tuning. 

Perhaps one of the unique ways to naturally address all of these problems is the extension of the MSSM to the $E_6$SSM{{\cite{Athron:2011ew}}. In such 
a scenario, not only is the GUT group enlarged beyond $SU(5)$ or $SO(10)$ to the exceptional group $E_6$, but the low energy (\ie, $\sim$ TeV 
scale) matter spectrum of the theory is also substantially modified by the presence of additional   
vector-like (with respect to the SM) superfields which are necessary to achieve complete anomaly cancellation, maintain coupling unification and to 
fill out full $E_6$ {\bf 27} fundamental representations. (It is important to note that the conventional two Higgs doublets responsible for the 
breaking of the usual SM 
gauge symmetries are also present at the TeV scale as in the MSSM.)  Furthermore, the $SU(2)_L \times U(1)_Y$ electroweak gauge structure of the SM is 
also likely to be extended at the TeV scale by additional $SU(2)$ and/or $U(1)$ factors{\cite{Hewett:1988xc}}. The simplest low-energy gauge 
extension from $E_6$ grand unification that is usually considered in the literature is that of a single new $U(1)$ factor. Such a $U(1)$ can 
result from a symmetry breaking chain such as: $E_6 \to SO(10)\times U(1)_\psi \to SU(5)\times U(1)_\chi \times U(1)_\psi \to SM \times U(1)_\theta$ 
where the $U(1)_\theta$ is a shorthand notation for some {\it a priori} arbitrary linear combination of both the $U(1)_\chi$ and $U(1)_\psi$ which 
occurs due to mass mixing.  
The heavy neutral gauge boson associated with this new gauge group, $Z'_\theta$, is very often discussed as a benchmark scenario 
and has been searched for (so far in vain) 
at the LHC{\cite {lhcsearches}} with lower limits on its mass now in excess of roughly $\sim $1.5 TeV. Of course, given the usual discussion of such 
possible symmetry breaking patterns within $E_6$, it is just as (perhaps more than?) likely that the two new neutral gauge bosons associated with 
both linear combinations of these new $U(1)$'s may survive down to the TeV scale. The $U(1)_\theta$ scenario can then be considered just a limiting 
case of this somewhat more complex situation which is mainly discussed only for its relative simplicity.    

If one wants to perform a detail examination of the interactions of both of these two new gauge bosons with the matter sector of the $E_6$SSM 
scenario at the TeV scale, the effects of 
gauge kinetic mixing (GKM) between the various $U(1)$'s can be of significant importance and cannot be neglected. 
In fact, the possible influence GKM on the properties and interactions of $Z'$'s, and the $Z'_\theta$ scenario in particular, were studied in some 
detail long ago{\cite {GKM}}. Much of this work focused on the issue of whether or not GKM can lead to near leptophobic $Z'_\theta$ couplings which 
would make such a state very difficult to observe at both the Tevatron and the LHC. In such an analysis, apart from an overall normalization factor 
of order unity, one finds that the $Z'_\theta$ couplings to fermions will now depend upon two parameters: the, in principle arbitrary, amount 
of $\chi-\psi$ mass mixing, expressed via the usual angle $\theta$, and the size of the GKM between the SM $U(1)_Y$ and $U(1)_\theta$ gauge fields, 
described by an additional parameter $\delta$, which itself is $\theta$-dependent. Both the overall strength factor as well as the parameter 
$\delta$ were shown to be uniquely 
calculable from a renormalization group equation (RGE) analysis once the TeV-scale matter content of the theory was specified. In this simplified 
case, it was found that{\cite {GKM}} exact leptophobia would not occur although a potentially significant cross section reduction for $Z'_\theta$ 
production in the usual dilepton channel might be possible.  In the analysis presented here we will return to this important issue in a different 
context where all three $U(1)$ factors are simultaneously present and their associated gauge fields undergo TeV-scale GKM. 

As is well-known, a necessary condition 
for GKM to occur in a (SUSY)GUT-based scenario, is the existence of incomplete GUT representations of matter superfields below the 
unification/high mass scale. This situation necessarily occurs within the standard $E_6$SSM scenario as, in addition to the three generations 
of {\bf 27} matter superfields at the TeV scale, an additional pair of Higgs doublets superfields (\ie, the `light' parts of the usual 
${\bf 5}+\overline{\mbox{\bf 5}}$ in the standard 
$SU(5)$ language), which are responsible for the breaking of the conventional SM gauge symmetries, are also necessarily present. Furthermore, at least 
in principle, additional light vector-like fields which form effective ${\bf 5}+\overline{\mbox{\bf 5}}$  
and/or ${\bf 10}+\overline{\mbox{\bf 10}}$ $SU(5)$ representations (but which are {\it incomplete} $E_6$ 
representations) may also be present without the loss of perturbativity and without disturbing the one-loop unification of couplings. In such 
a scenario, GKM, though absent by definition at the high mass scale due to the presence of complete GUT representations, is then generated 
radiatively at the TeV scale via the RGEs from the `non-orthogonality' of the full set of the various 
$U(1)$ gauge charges. Thus, once the full low energy matter content of the theory is specified, one can use the RGEs to determine all of the 
TeV-scale gauge couplings as well as all of the GKM parameters, at least numerically, with the mixing angle $\theta$ remaining as the only 
undetermined free parameter. 

As mentioned above, in this paper we will consider the situation where both of the two new $U(1)$'s are present at the TeV scale in the $E_6$SSM 
framework such that GKM can (and must) occur {\it simultaneously} between the gauge fields associated with all three $U(1)$ group factors: 
$U(1)_Y, U(1)_\chi$ and $U(1)_\psi$. Note that a detailed discussion of the appropriate formalism for GKM among three general $U(1)$ factors has been 
recently given in Ref.{\cite{Heeck:2011md}}; we will make direct use of the results presented there in the analysis that follows.{\footnote {See, 
however, Ref.{\cite{Braam:2011xh}} for a different approach wherein the authors seek a possible basis change in the group space of the multiple 
$U(1)$ factors to remove GKM completely at the one-loop level.}} As we will 
see below, in such a situation, the coupling structure of both of the new gauge $Z'$ bosons are now 
potentially more complex than in the simpler case of the $U(1)_\theta$ model and will be determined in terms of five 
`Lagrangian'  parameters (apart from the common overall strength factor): the usual $\chi-\psi$ mixing angle, $\theta$, three GKM 
factors, which we will call $\delta_{1,2,3}$, as well as the relative magnitudes of the $g_\chi$ and $g_\psi$ gauge couplings at the TeV scale. 
Note that since we now work in the $\chi-\psi$ basis, unlike in the $U(1)_\theta$ model, none of these parameters are $\theta$-dependent. 
Apart from $\theta$ itself, all of these parameters, as well as the overall strength factor will be shown to be calculable within an extended 
version of the previous RGE analysis. Furthermore, due to manner in which the various fermion couplings to the $Z'$ fields can be expressed in terms 
of these calculable Lagrangian parameters, we will show that the actual number of {\it independent} parameters is actually only three. 

The outline of this paper is as follows: Section 2 contains an analysis and discussion of the various couplings and relevant RGE machinery 
necessary to calculate the TeV-scale values of the various parameters listed above. In Section 3 we discuss the constraints on $E_6$SSM models 
with additional matter fields arising from split supermultiplets and construct a sizeable set of $E_6$SSM models which may potentially lead to 
significant GKM effects. The five coupling parameters apart from $\theta$ are then numerically determined for this set of models and are compared 
and contrasted in Section 4. The extension of this analysis to model sets which also include additional SM singlet fields is also considered briefly 
in this Section as is an analysis of some of the phenomenological implications of general GKM. An overall summary of our results and conclusions 
are to be found in Section 5.

\section{Analysis I: RGEs}

We begin our analysis by reviewing the formulation of gauge kinetic mixing between three $U(1)$ factors as given in{\cite{Heeck:2011md}}, realizing 
it within the specific context of the $E_6$SSM scenario. 
Consider the Lagrangian for the electroweak part of the SM with the addition of two new $U(1)$ fields which is decomposed in the following manner:
\begin{equation}
{\cal L}={\cal L}_{kin}+{\cal L}_{int}+{\cal L}_{SSB}+{\cal L}_{SUSY}\,.
\end{equation}
In the absence of spontaneous symmetry breaking at the TeV scale, the general form of the Lagrangian for the pure electroweak gauge kinetic piece 
in the $E_6$SSM scenario can be written as 
\begin{eqnarray}
{\cal L}_{kin}=-{1\over {4}}W^i_{\mu\nu}W^{i\mu\nu}-{1\over {4}}\tilde B^{\mu\nu}\tilde B_{\mu\nu}-{1\over {4}}
\tilde Z^{1\mu\nu}\tilde Z_{1\mu\nu}-{1\over {4}}\tilde Z^{2\mu\nu}\tilde Z_{2\mu\nu}\,,\nonumber\\
-{\sin \alpha \over {2}}\tilde B^{\mu\nu}\tilde Z_{1\mu\nu}-{\sin \beta \over {2}}\tilde B^{\mu\nu}\tilde Z_{2\mu\nu}
-{\sin \gamma \over {2}}\tilde Z_1^{\mu\nu}\tilde Z_{2\mu\nu}\,,
\end{eqnarray}
with $W^i,\tilde B$ and $\tilde Z_{1,2}$ representing the usual $SU(2)_L$, $U(1)_Y$ and the two 
new $U(1)_{1,2}$ fields (which we will later identify with the $Z_{\chi,\psi}$ gauge bosons), respectively,  
with the index `i' labeling the weak isospin. Note that the second line above contains the form for the general gauge kinetic mixing allowed 
among the three $U(1)$ fields. In this basis the generic gauge interaction terms for the various fermions can be written as
\begin{equation}
{\cal L}_{int}=-\bar \psi \gamma^\mu[g_LT^iW^i_\mu+\tilde g_Y Y \tilde B_\mu+\tilde g_1 Q_1\tilde Z_{1\mu} +(1\to 2)]\psi\,,
\end{equation}
where $\tilde g_{1,2}, Q_{1,2}$ are the couplings and appropriate group generators, respectively, corresponding to the new fields 
$\tilde Z_{1,2\mu}$. 

As was shown in Ref.{\cite{Heeck:2011md}}, these  
kinetic terms can be brought into a diagonal canonical form by a suitable set of field redefinitions: $(\tilde B, \tilde Z_1, \tilde Z_2)^T=
U (B,Z_1,Z_2)^T$, 
where $U$ is (effectively) a triangular matrix as given in full detail in {\cite{Heeck:2011md}}. In terms of the elements of the matrix $U$, this 
transformation takes the 
explicit form (note that it is always true that $U_{11}=1$) 
\begin{eqnarray}
\tilde B &=& B+U_{12} Z_1+ U_{13} Z_2\,, \nonumber \\
\tilde Z_1 &=& U_{22} Z_1 +U_{23} Z_2\,, \nonumber \\
\tilde Z_2 &=& U_{33} Z_2\,.
\end{eqnarray}
With the corresponding rescalings of the original gauge couplings 
\begin{eqnarray}
g_Y &=& \tilde g_Y\,, \nonumber \\
g_1 &=& \tilde g_1 U_{22}\,, \nonumber \\
g_2 &=& \tilde g_2 U_{33}\,, \nonumber \\
g_{Y1} &=& \tilde g_Y U_{12}\,,\nonumber \\
g_{Y2} &=& \tilde g_Y U_{13}\,,\nonumber \\
g_{12} &=& \tilde g_1 U_{23}\,,
\end{eqnarray}
the fermionic interactions with the gauge fields can be written in a more familiar form as
\begin{equation}
{\cal L}_{int}=-\bar \psi \gamma^\mu[g_LT^aW^a_\mu+g_Y Y B_\mu+(g_1 Q_1+g_{Y1} Y) Z_{1\mu} +(g_2 Q_2+g_{Y2} Y +g_{12} Q_1) Z_{2\mu}]\psi\,.
\end{equation}
Symbolically, for later ease in the analysis below and to make contact with the notation employed earlier in the last paper of Ref.{\cite {GKM}}, 
the abelian 
$U(1)_Y\times U(1)_1 \times U(1)_2$ part of the term in the square bracket above can be re-written into a slightly more generic looking form, 
dropping the Lorentz index,  as 
\begin{equation}
{\cal L}_{abelian} \sim g_{aa} Q_a Z_a+(g_{bb} Q_b+g_{ab} Q_a) Z_b  +(g_{cc} Q_c+g_{ac} Q_a +g_{bc} Q_b) Z_c\,,
\end{equation}
(with no summation of repeated indices implied) or, equivalently 
\begin{equation}
{\cal L}_{abelian}\sim g_{aa} Q_a Z_a+ g_{bb}(Q_b+\delta_1 Q_a) Z_b  +g_{cc} (Q_c+\delta_2 Q_a +\delta_3 Q_b) Z_c\,,
\end{equation}
where here we have defined the GKM parameters $\delta_1=g_{ab}/g_{bb}$ \etc. Note that later we will identify the indices (a-c) with our specific 
set of $U(1)$'s as $a=Y$, $b=\chi$ 
and $c=\psi$ but we will sometimes find it convenient so use both sets of notation simultaneously.  Finally, by normalizing the gauge couplings to 
that of the SM hypercharge, this interaction can also be written as
\begin{equation}
{\cal L}_{abelian}\sim g_{aa} [Q_a Z_a+ \lambda_b(Q_b+\delta_1 Q_a) Z_b  +\lambda_c (Q_c+\delta_2 Q_a +\delta_3 Q_b) Z_c]\,,
\end{equation}
where $\lambda_{b,c}= g_{bb,cc}/g_{aa}$, \ie, are the relative strengths of the new $U(1)$ couplings in terms of that for SM hypercharge. 
The size of the various $\delta_i$ 
not only reflect the magnitude of the off-diagonal gauge couplings 
in the equation above but also the size of the kinetic mixing in the original Lagrangian, \ie, they will be zero when $\sin \alpha$, \etc, 
are zero. As discussed 
above, since kinetic mixing is assumed to be absent at the high mass scale due to the presence of compete GUT representations, the resulting 
values for the 
$\delta_i$ at the TeV scale will only be generated by RGE running in the presence of the incomplete {\bf 27} representations at the 
TeV scale. As we 
will see below, given some knowledge of what these representations can be, the values of $\delta_i$, as well as the various gauge couplings 
themselves, become 
calculable (at least numerically) by solving the corresponding RGE equations. Note that although the number of parameters appearing in 
${\cal L}_{abelian}$ is large only certain fixed combinations will appear in the fermion couplings themselves. This will be discussed further below. 

Note that in what follows the various `diagonal' gauge couplings, $g_{aa}$ \etc, will be taken to be `GUT' normalized in this basis since 
we will assume that 
complete $E_6$ representations exist at the high mass scale. Thus we will employ the SM notation and normalization conventions, \ie, $Y\to 
\sqrt {3\over {5}} Y_{SM}$ and 
$\tilde g_Y\to \sqrt {5\over {3}} g'$ such that $Q_{em}=T_{3L}+Y_{SM}$. Furthermore, for the specific case at hand, one also finds that $Q_1 
= Q_\chi/2\sqrt 10$ 
and $Q_2= Q_\psi/2\sqrt 6$ employing this standard normalization convention. For completeness, the values of the constants $Q_{\chi,\psi}$ 
for the various fermions in the $E_6$ {\bf 27} representation, employing the notation of the last paper in 
Ref.{\cite{Hewett:1988xc}, are given in Table~\ref{numbers}.

%%%%% Another Table
%%
\begin{table*}[htbp]
\leavevmode
\begin{center}
\label{numbers}
\begin{tabular}{lccccc}
\hline
\hline
Particle & $SU(3)_c$ & $2\sqrt{6} Q_\psi$ & $2\sqrt{10} Q_\chi$ &
$2\sqrt{15} Q_\eta$ & Y \\
\hline
$Q=(u,d)^T$   & {\bf 3}      &   1   & -1   & 2 & 1/6    \\
$L=(\nu,e)^T$ & {\bf 1}      &   1   &  3   & -1 &-1/2   \\
$u^c$ &$\overline{\mbox{\bf 3}}$       &   1   &  -1  & 2& -2/3   \\
$d^c$ &$\overline{\mbox{\bf 3}}$       &   1   &  3   & -1 &1/3    \\
$e^c$         &{\bf 1}       &   1   &  -1  &  2 &1     \\
$\nu^c$       &{\bf 1}       &   1   &  -5  &  5 &0     \\
$H=(N,E)^T$   &{\bf 1}       &  -2   &  -2  &  -1 &-1/2  \\
$H^c=(N^c,E^c)^T$ &{\bf 1}   &  -2   &  2   &  -4 &1/2   \\
$h$           & {\bf 3}      &  -2   &  2   &  -4 &-1/3  \\
$h^c$ &$\overline{\mbox{\bf 3}}$       &  -2   &  -2  & -1 &   1/3  \\
$S^c$     &{\bf 1}           &   4   &  0   &  5 &0     \\
\hline
\hline
\end{tabular}
\caption{Quantum numbers of the particles contained in the {\bf 27} representation of $E_6$ in the notation of the last paper in 
Ref.{\cite{Hewett:1988xc}}; standard particle embeddings are assumed and all fields are taken to be left-handed.}
\end{center}
\end{table*}

In order to constrain the low scale values of the $\lambda_{\psi,\chi}$ and $\delta_{1-3}$ for any given model we must first perform an RGE 
analysis. 
At one-loop the RGE equations for the usual $SU(3)_c$ and $SU(2)_L$ gauge couplings are decoupled from those in the $U(1)$ sector so that the 
associated RGEs can be
trivially analytically integrated. Writing $L=\log (M_U/M_Z)$, these two RGE equations can be combined in the usual manner from which we 
obtain the standard expressions
\begin{eqnarray}
L &=& {{2\pi (\alpha^{-1}_s-x_w\alpha^{-1})}\over {\beta_s-\beta_L}}
\,, \nonumber \\
\alpha^{-1}_U  &=&  {{\beta_s x_w \alpha^{-1}-\beta_L\alpha^{-1}_s}
\over {\beta_s-\beta_L}} \,,
\end{eqnarray}
with $\alpha_U$ being the common unification coupling and $x_w=\sin^2 \theta_w$. For numerical purposes we will take $\alpha_s(M_Z)=0.1186$, 
$\alpha_{em}^{-1}(M_Z)=127.957$  and $\sin ^2 \theta_w=0.2315$ in the analysis below; our results will depend only very weakly on these particular 
choices.

Defining $B_{ij}= Tr (Q_iQ_j)$ where $Q_{i=a,b,c}$ is defined from the Lagrangian above and the trace is taken over the fields in the 
theory below the GUT scale we can then consider the RGE equations for the various couplings. 
For the $U(1)_{Y,\chi}$ couplings and for the parameter $\lambda_\chi$ we can proceed exactly as in the simpler case of only two $U(1)$'s 
since the RGE for $U(1)_\psi$ decouples at this order. Most easily obtained are the first set of partially coupled RGEs: 
\begin{eqnarray}
{dg_{aa}\over {dt}} &=& {g^3_{aa}\over {16\pi^2}}B_{aa}\,, \nonumber \\
{dg_{bb}\over {dt}} &=& {g^3_{bb}\over {16\pi^2}}[B_{bb}+\delta_1^2 B_{aa}+2\delta_1 B_{ab}]\,, \\
{dg_{ab}\over {dt}} &=& {1\over {16\pi^2}}[2g^2_{aa}g_{bb}B_{ab}+2g^2_{aa}g_{ab}B_{aa}+g_{ab}g^2_{bb}B_{bb}+g^3_{ab}B_{aa}+2g_{bb}g^2_{ab}B_{ab}]\,, 
\nonumber
\end{eqnarray}
Integration of the first RGE above yields the standard result for the hypercharge coupling (recalling that $a=Y$): 
\begin{equation}
g^{-2}_{aa=Y}(t)={\alpha^{-1}_U\over {4\pi}}[1+{\alpha_U \over {2\pi}} B_{aa(=YY)}(t_U-t)]\,,
\end{equation}
where $t_U \sim \log M_U$ and the standard unification boundary condition has been imposed at the high scale.
Since $\delta_1=g_{ab}/g_{aa}$, and the solution for $g^2_{aa}(t)$ is known, we can combine the last two of the RGEs above to obtain
\begin{eqnarray}
{d\delta_1\over {dt}}&=&{1\over {g_{bb}}}{dg_{ab}\over {dt}}-
{g_{ab}\over {g^2_{bb}}}{dg_{bb}\over {dt}}\,, \nonumber \\
     &=& {g^2_{aa}\over {8\pi^2}}[B_{ab}+\delta_1 B_{aa}]\,.
\end{eqnarray}
This can now be directly integrated yielding the earlier obtained, well-known result{\cite {GKM}}
\begin{equation}
\delta_1(t)=-{B_{ab}\over {B_{aa}}}\Big[1-[1+{\alpha_U B_{aa}(t_U-t)\over {2\pi}}]^{-1}\Big]\,,
\end{equation}
where we have imposed the boundary condition that $\delta_1(t)$ vanishes at the high mass scale $M_U$ since we have assumed that complete
multiplets exist there. The weak/TeV scale parameter $\delta_1$ relevant for the $Z_{1,2}$ couplings is obtained when we set $t\sim \log M_Z$ so that
$t_U-t=\log (M_U/M_Z)=L$ in the expression above. Note that $\delta_1$ grows linearly as the value of $B_{ab}$ is increased. 

Given the analytic result for $\delta_1(t)$ then allows us to re-write the RGE for $g^2_{bb}$ as
\begin{equation}
{dg_{bb}\over {dt}} = {g^3_{bb}\over {16\pi^2}}[B_{bb}+2\delta_1(t)B_{ab}+\delta^2_1(t)B_{aa}]\,,
\end{equation}
which also can then be analytically integrated. Defining the parameter combination $z=\alpha_UB_{aa}L/2\pi$ we obtain as before 
\begin{equation}
g^{-2}_{bb}(M_Z)={\alpha^{-1}_U\over {4\pi}}+{B_{bb}L\over {8\pi^2}}\Big[1-{B^2_{ab}\over {B_{aa}B_{bb}}}{z\over {1+z}}\Big]\,. 
\end{equation}
Recalling that $b=\chi$ for our setup of interest, this result also allows us to obtain the numerical value for the ratio of couplings 
$\lambda_{\chi}= g_{bb}/g_{aa}$. Continuing on to the other parameters straightforwardly leads us to the following more highly coupled set of RGEs: 
\begin{eqnarray}
{dg_{cc}\over {dt}} &=& {g^3_{cc}\over {16\pi^2}}[B_{cc}+\delta_2^2 B_{aa}+\delta_3^2 B_{bb}+2\delta_2 B_{ac}+2\delta_3 B_{bc}+2\delta_2 
\delta_3 B_{aa}]\,,  
\nonumber \\
{d\delta_{3}\over {dt}} &=& {g^2_{bb}\over {8\pi^2}}[B_{bc}+\delta_2B_{ab}+\delta_3 B_{bb}+\delta_1 B_{ac}+\delta_1\delta_2 B_{aa}+\delta_1 
\delta_3 B_{ab}]\,, \\
{d\delta_{2}\over {dt}} &=& {g^2_{aa}\over {8\pi^2}}[B_{ac}+\delta_2B_{aa}+\delta_3 B_{ab}]+{g^2_{bb}\over {8\pi^2}}\delta_1 \Big [[B_{bc}+
\delta_2 B_{ab}+
\delta_3 B_{bb}]\nonumber \\
 & & +\delta_1[B_{ac}+\delta_2 B_{aa}+\delta_3 B_{ab}]\Big]\,.\nonumber 
\end{eqnarray}
which, unlike those discussed above, can only be solved by employing numerical methods. This we do by making use of the previously obtained 
analytical results as inputs. In order to proceed further in such a numerical analysis, however, we must also obtain a set of values for the 
$B_{ij}$ coefficients which appear in the equations above. These depend on the details of the matter superfield content of the low-energy theory 
to which we must now turn.

\section{Analysis II: Sample Models}

As discussed above, GKM will only occur if a set of incomplete $E_6$ multiplets is present at the $\sim$ TeV scale and so we must determine what a 
realistic set of possibly low-energy spectra might be if we want to estimate the magnitude of GKM effects. This was discussed at some length in 
earlier work{\cite {GKM}} so here we provide just a summary of and short elaboration 
on that detailed analysis. 

The set of possible low-energy fields directly follows from a short list of model building requirements:
\begin{itemize}

\item  The SM gauge couplings and those of the new $U(1)$'s are assumed to unify at a high scale as in the MSSM and we don't want this important  
feature to be disturbed. This implies that: ($i$) as far as particles which 
carry SM quantum numbers are concerned we can add only sets of particles that would normally form complete multiplets under $SU(5)$.  ($ii$) The 
number and types of 
these new fields is clearly restricted since perturbative unification is lost if too many multiplets are added. ($iii$) A priori, the number 
of SM singlet fields 
is not so easily restricted.  

\item  SM gauge anomalies must cancel amongst the low energy matter fields in the model. This suggests that only fields which are vector-like 
or singlets with respect to the SM gauge group be considered. In addition, the anomaly freedom must be maintained when the two new $U(1)$ 
fields are added to the MSSM. Employing only singlets/vector-like fields also assists with satisfying constraints arising from precision 
measurements.  

\item  When taken together the above requirements strongly suggests that at low energies we can augment the particle spectrum by at most four 
complete ${\bf 5}+\overline{\mbox{\bf 5}}$'s or only one ${\bf 5}+\overline{\mbox{\bf 5}}$ plus one ${\bf 10}+\overline{\mbox{\bf 10}}$ 
(in addition to possible $SU(5)$ singlets) beyond that of the usual MSSM. Note that some of this additional field content, three generations 
of  ${\bf 5}+\overline{\mbox{\bf 5}}$'s plus singlets, {\it already} appears in the three ${\bf 27}$'s present in the $E_6$SSM. Thus we can only 
add, at most, one set of ${\bf 5}+\overline{\mbox{\bf 5}}$ fields plus singlets to the low energy content of the {\it minimal} $E_6$SSM scenario.  

\item  The new matter fields will be assumed to originate from either ${\bf 27}+\overline{\mbox{\bf 27}}$ combinations or from {\bf 78}'s of $E_6$ 
since these will automatically be both vector-like as well as anomaly free under the full $E_6$ group. 

\end{itemize}

With this set of requirements there are actually only a small number of potentially interesting specific cases to consider.  In the minimal 
$E_6$SSM we know that the low energy theory already contains three complete {\bf 27}'s as well as the usual pair of Higgs doublets (here labeled 
as $H_1,H_1^c$). These Higgs fields are then easily identified as the minimal split multiplet content at low energies. The specific choice of 
origin of these Higgs within the ${\bf 27}+\overline{\mbox{\bf 27}}$ or the ${\bf 78}$ 
will then tell us how they transform with respect to the two new $U(1)$ gauge groups. As noted above, 
since the three {\bf 27}'s already contain three pairs of 
${\bf 5}+\overline{\mbox{\bf 5}}$ fields beyond the usual MSSM, this implies that we are only free to add a single extra ${\bf 5}+
\overline{\mbox{\bf 5}}$ 
(plus possible singlets) to the low energy spectrum of the $E_6$SSM without upsetting unification. Here we note that ${\bf 27}+
\overline{\mbox{\bf 27}}$'s contain 
three different possible embeddings of ${\bf 5}+\overline{\mbox{\bf 5}}$: (1)~${\bf 5}(-2,2)+\overline{\mbox{\bf 5}}(2,-2)$, (2)~${\bf 5}(2,2)
+\overline{\mbox{\bf 5}}(-2,-2)$ and (3)~${\bf 5}(-1,-3)+\overline{\mbox{\bf 5}}(1,3)$, where the numbers in the parentheses designate 
the $Q_{\psi,\chi}$ quantum 
numbers as normalized in Table 1. However, the {\bf 78} contains only a lone possibility: (4)~${\bf 5}(3,-3)+\overline{\mbox{\bf 5}}(-3,3)$. 
Given this information we can calculate any and all possible contributions to the various $B_{ij}$'s appearing in the equations above, as well as 
those to $\beta_{s,L}$, in an unambiguous manner. 

What about SM singlet fields?  Within the ${\bf 27}+\overline{\mbox{\bf 27}}$, we already have the fields $S^c$ and $\nu^c$ (plus their conjugates) 
that transform 
as ${\bf 1}(4,0)+{\bf 1}(-4,0)$ and ${\bf 1}(1,-5)+{\bf 1}(-1,5)$, respectively. On the otherhand, the ${\bf 78}$ contains only one possibility, 
here simply denoted as $X$ (plus its 
conjugate), which transform as ${\bf 1}(-3,-5)+{\bf 1}(3,5)$. Note that since they are SM singlets, the presence of these additional fields 
will only alter the 
values of $B_{\chi \chi},B_{\chi \psi}$ and $B_{\psi \psi}$. While the set of possible additional TeV-scale SM non-singlet fields is rather 
restricted, it is 
possible that a sizeable and a priori unknown number of these new SM singlet fields may also be present.  To be concrete in the analysis 
that follows we will mostly  
ignore the possible addition of SM singlets. We have verified, however, that the addition of one or two complete sets of the above singlets in 
various combinations will only alter our numerical 
results for the values of the $\delta_i$ in the models considered below at the level of at most $\sim 10-20\%$. 
However, the possible changes to both 
$\lambda_{\chi,\psi}$ are found to be potentially more significant, up to $\sim 30-35\%$. This is to be expected as the most direct effect of the 
additional singlets, even in 
the absence of GKM, would be to reduce the values of the $g_{\chi,\psi}$ couplings due to the enhanced RGE running from the high scale. We 
will return to a discussion of the possible influence of additional SM singlet fields on our results in the analysis below. 

From the discussion above, we can see that as far as SM non-singlet fields are concerned there are only two possible subcases to consider. 
Either ($i$) $H_1/H_1^c$ is the only pair of light superfields beyond the three {\bf 27}'s or ($ii$) one additional complete set of ${\bf 5}+
\overline{\mbox{\bf 5}}$ 
fields is also present. In case ($i$), with only 4 choices for the $H_1/H_1^c$ quantum numbers, the calculation is very straightforward. 
However,in case ($ii$) 
the situation is somewhat more complex since we can think of the additional low energy spectrum as containing two `Higgs' doublets, 
$H_{1,2}/H_{1,2}^c$ as well 
as a pair of isosinglet, color triplet superfields, $D_1,D_1^c$ which fill out the remainder of the full ${\bf 5}+\overline{\mbox{\bf 5}}$. 
This field content now allows for $4^3=64$ possible(but not necessarily independent) quantum 
number assignments leading to different values for $B_{ij}$'s. We label these possibilities by the set of integers (i,j,k) where the 
first(second,third) index labels the 
embedding choice, \ie, (1)-(4), for the field $H_1/H_1^c$($H_2/H_2^c$,~$D_1/D_1^c$). For example, we may choose $H_1/H_1^c$ to be from (1), 
$H_2/H_2^c$ from (3) 
and $D_1/D_1^c$ from (4) and we would label this particular subcase as (1,3,4). The calculation is now again straightforward. Thus there are now 
a total of 68 toy model scenarios (without SM singlets) to consider.

\section{Analysis III: Numerical Results}

The first step in the examination of the $Z'$ couplings is to obtain the complete set of values for the 5 parameters $\delta_{1,2,3}$ and 
$\lambda_{\chi,\psi}$; these 
results are shown in Figs.~\ref{fig1} and \ref{fig2} for all of the 68 models discussed in the previous section.  In each of these plots we 
see a discrete set of model points. Note that in some cases two (or more) of the 68 models may lead to identical `degenerate' values for any 
given pair of the parameters that are displayed here so that several of the locations may actually be multiply occupied. Note that in all 
cases the ranges 
of the $\delta_i$ away from 0 are not very large nor are the deviations of $\lambda_{\chi,\psi}$ from unity. As mentioned 
above, the addition of one or two sets of $SU(5)$ singlet fields of various types is found to lead to possible upward shifts in the magnitudes of the 
$\delta_i$ shown here by less than $\simeq 15-20\%$ while the corresponding shifts in the $\lambda_i$ (generally to smaller values) can be as large 
as $\simeq 30-35\%$. Of course adding further singlets will enlarge these effects.

\begin{figure}[htbp]
\centerline{
\includegraphics[width=8.0cm,angle=90]{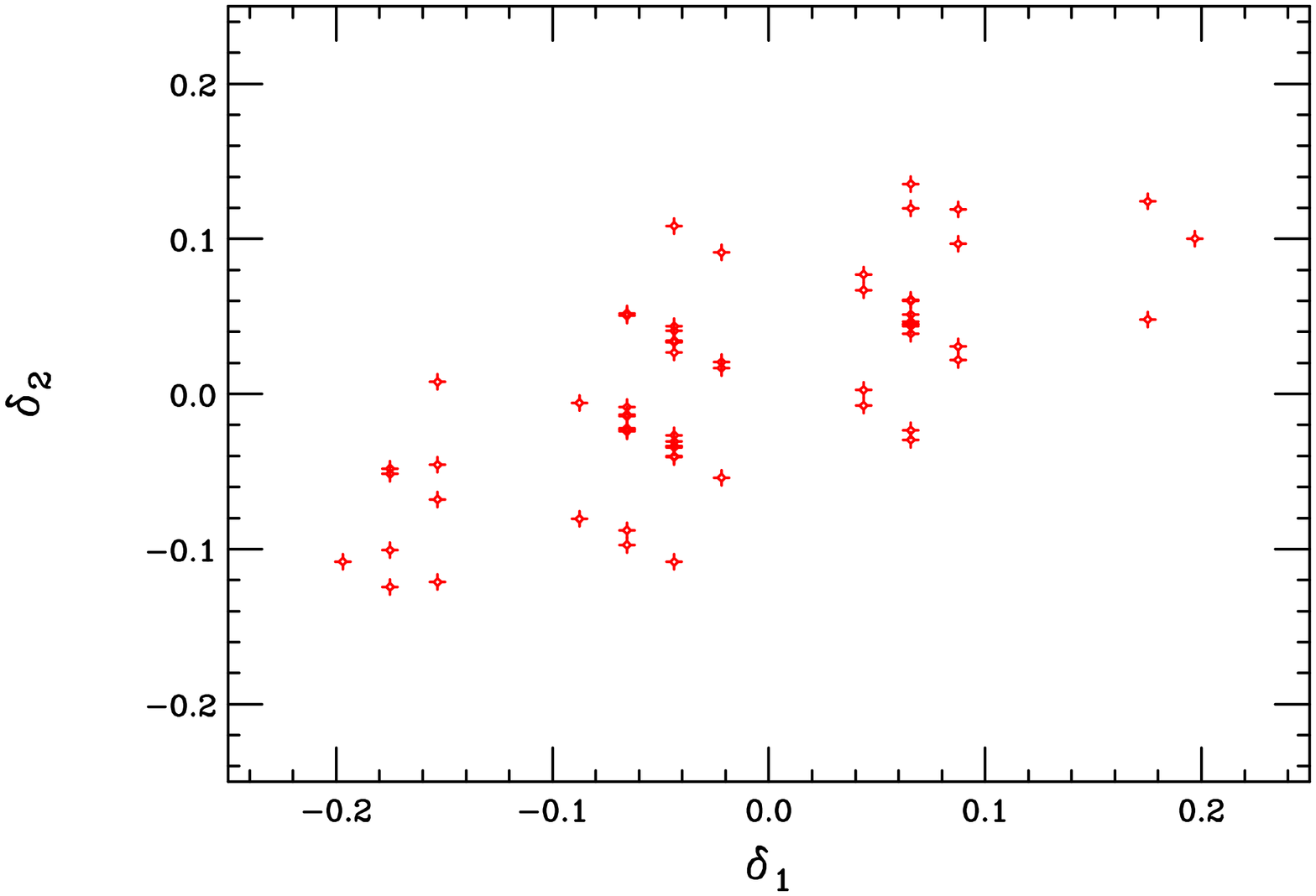}}
\vspace*{0.5cm}
\centerline{
\includegraphics[width=8.0cm,angle=90]{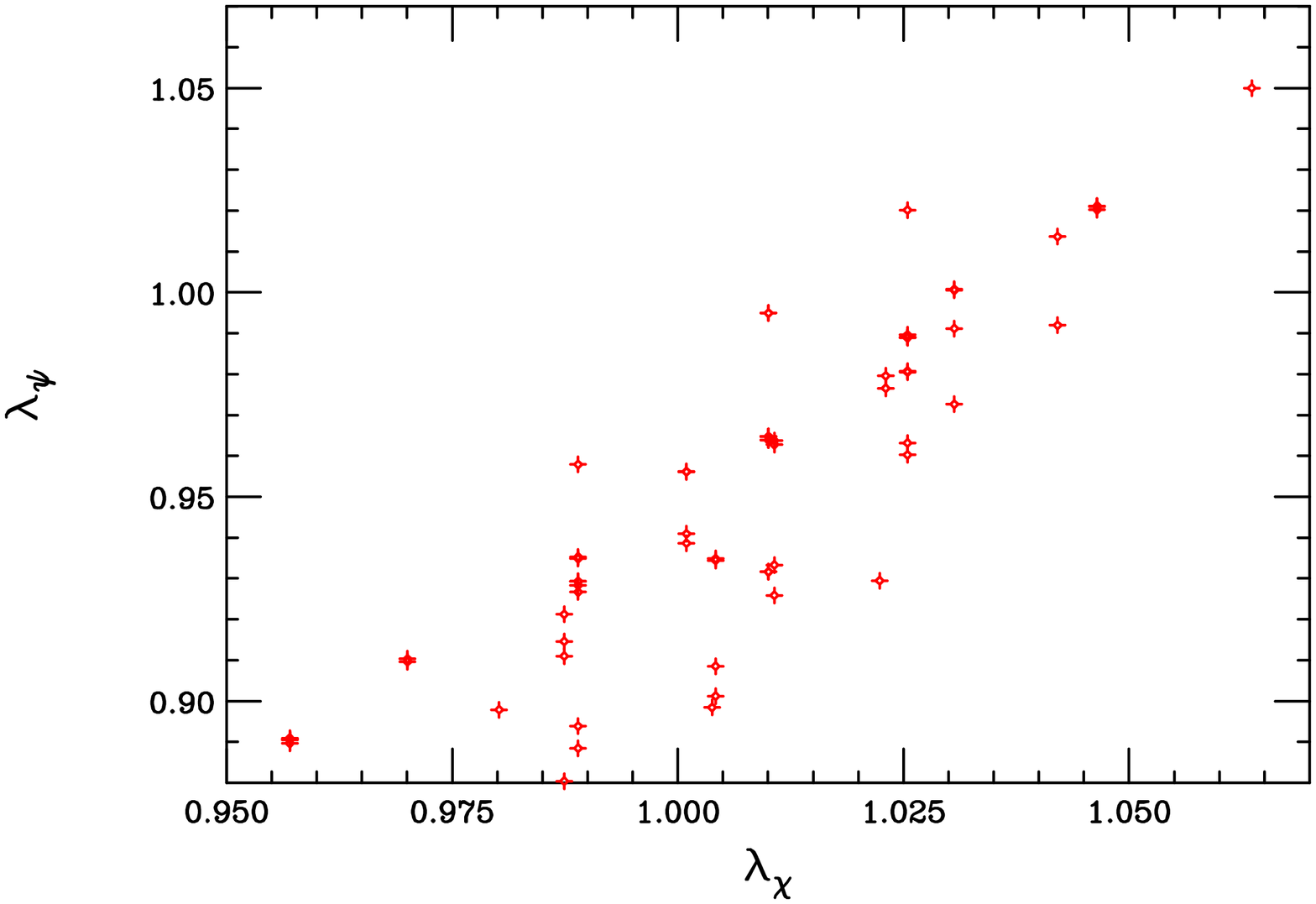}} 
\vspace*{0.5cm}
\caption{Values of (bottom) $\lambda_{\chi,\psi}$ and (top) $\delta_{1,2}$ for the set of 68 models discussed in the text.}
\label{fig1}
\end{figure}
\begin{figure}[htbp]
\centerline{
\includegraphics[width=8.0cm,angle=90]{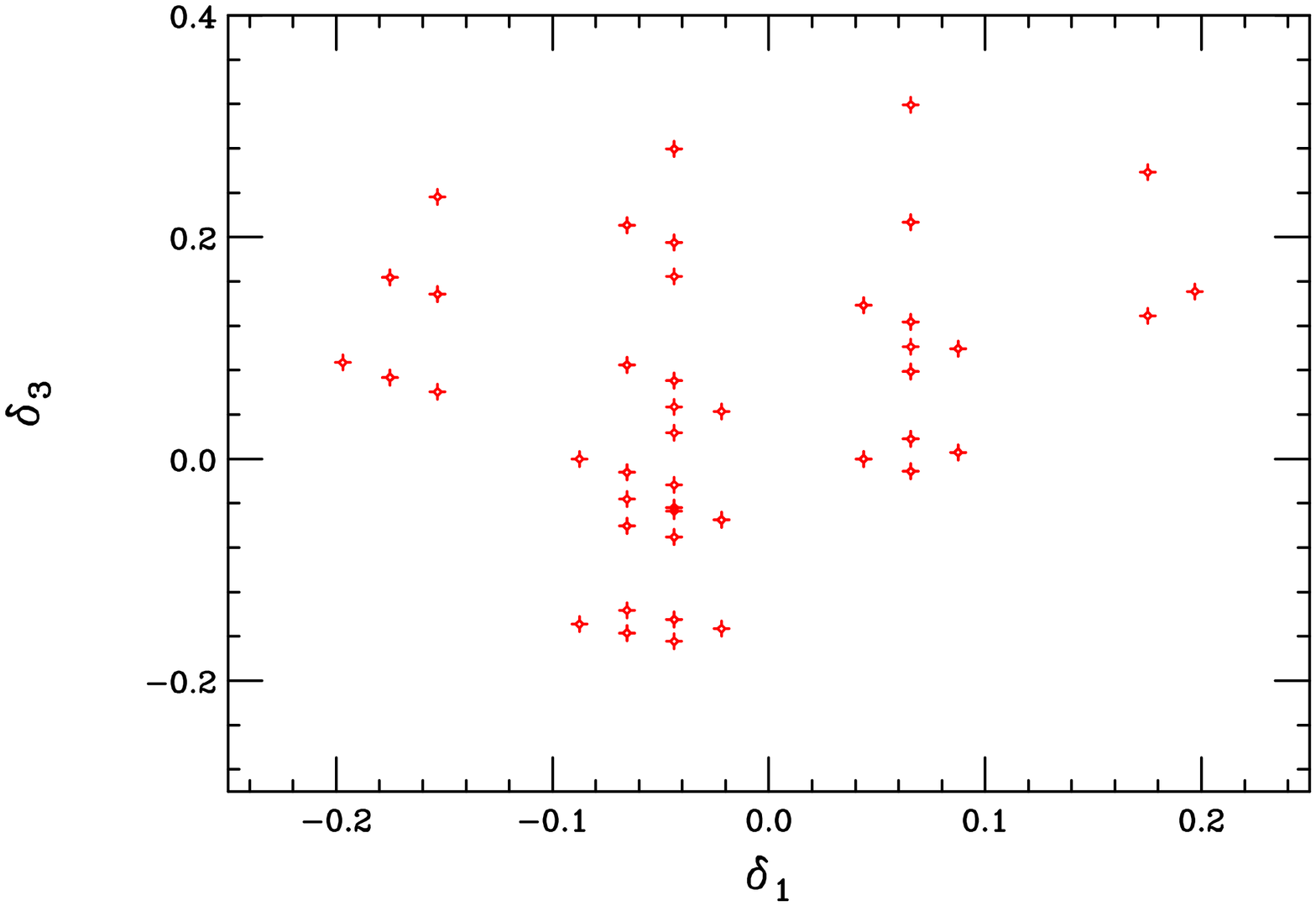}}
\vspace*{0.5cm}
\caption{Same as the previous figure but now showing the results for $\delta_{1,3}$}
\label{fig2}
\end{figure}

Let us now turn to the predictions for the couplings of the $Z'$ to SM fermions. One of issues that we can now immediately address is this  
framework is the question of whether or not these general GKM modifications to the fermion interactions can lead to a near leptophobic coupling for 
either of the new gauge bosons; as we noted above this subject has had a long history{\cite {GKM}}. As can be easily found in the $E_6$ model 
case, complete leptophobia certainly does not occur in the absence of GKM for any value of the parameter $\theta$ and, as noted above, does 
not occur for the $Z'_\theta$ model with GKM. The reason that this is an important issue is, of course, that the typical search channel for a 
new $Z'$ gauge boson at a hadron collider such as the LHC is the resonant production of Drell-Yan pairs. However, to determine whether or not the 
event rate in this channel is significantly altered, in the narrow width approximation (NWA), one needs to determine the product of the $Z'$ 
production 
cross section times the corresponding leptonic branching fraction, \ie, $\sigma_{Z'}\cdot B_{\ell=e,\mu}$.  Clearly, if the value of $B_\ell$ is 
reduced by a shift in the $Z'$ leptonic couplings this will make this new particle more difficult to discover. However, this reduction might be 
simultaneously compensated for in the overall $Z'$ production cross section. Thus leptophobia is not just an issue of reduced leptonic couplings. 
Using the NWA, the predicted dilepton event rate can be more fully expressed as $[(v_u^2+a_u^2)f_u +(v_d^2+a_d^2)f_d]B_\ell {\cal L}$, where 
${\cal L}$ is the collider integrated luminosity, $v_u(a_u)$ is the vector(axial-vector) coupling of the $u$ quark to the $Z'$ and $f_u$ is an 
integral over the appropriate product of the parton densities which only depends upon the $Z'$ mass (and similarly for the $d$ quark). 
{\footnote {It is interesting to recall that generation independence plus the assumption that the linear combination of group generators to which 
the $Z'$ couples (such as the $U(1)_{\chi,\psi}$ generators) commutes with weak isospin implies that all of the various SM fermion couplings 
can be expressed in terms of no more than 5 independent parameters.}}  
From this we see that actually all of the different $Z'$ couplings are involved in determining the overall event rate and not just the leptonic  
ones. Also, quite obviously a hadrophobic $Z'$ is just as bad an outcome as a leptophobic $Z'$ as far as this search channel is concerned. Clearly 
it is more important to consider how GKM affects the product $\sigma_{Z'}\cdot B_\ell$ rather than simply just considering $B_\ell$ alone if we are 
trying to understand any suppression of hadron collider $Z'$ event rates. 

Give this observation we can now directly address the question: `What is the influence of GKM on $Z'$ production rates for the general mixing case 
considered here?' 
For purposes of this discussion we will assume for simplicity that the lightest $Z'$ state does not have kinematically allowed decays into any 
of the non-SM fields (including both the additional $E_6$ SM singlets/vector-like exotics as well as any of the superpartners) and that this $Z'$ 
does not have any significant mass mixing with the SM $Z$ so that decays such as, \eg, $Z'\to W^+W^-$ can be safely ignored. 
To address the question of collider event rates it is sufficient to consider the various fermionic interactions of a single general linear 
combination of the $Z_{\chi,\psi}$ fields, the $Z_1'$, \ie, the lighter of the two mass eigenstates. In the notation used above the couplings 
of the $Z_1'$ can be explicitly written as   
\begin{equation}
{g\over {c_w}} \sqrt{{5x_w}\over {3}} \Bigg[-s\lambda_\chi {Q_\chi\over {2\sqrt {10}}} + c\lambda_\psi {Q_\psi\over {2\sqrt {6}}}+
\sqrt{3\over {5}} {Y\over {2}}
(-s\lambda_\chi \delta_1+c\lambda_\psi \delta_2)+c\lambda_\psi \delta_3 {Q_\chi\over {2\sqrt {10}}} \Bigg]\,,
\end{equation}
where $s(c)=\cos \theta_6 (\sin \theta_6)$ is the effective $Z_\chi-Z_\psi$ mixing angle. The corresponding couplings for the heavier $Z_2'$ are 
trivially obtainable from those above by the simple replacements $(-s,c)\to (c,s)$. Interestingly, we note that making the replacements 
$\lambda_{\chi,\psi} \to \lambda$, $\delta_1 \to -s\delta,~\delta_2 \to c\delta$ and $\delta_3 \to 0$ we recover the coupling structure 
associated with the earlier obtained GKM modifications to the $U(1)_\theta$ model{\cite {GKM}}. However, in that case it was seen that both 
the overall strength parameter, $\lambda$, as well as the GKM parameter, $\delta$, were both dependent on the value of $\theta$ which is not the 
situation in the triple GKM mixing scenario considered here. Note further that the `standard' result for the $Z'$ 
couplings that are most usually employed for $E_6$-type model studies is given solely by only the first two terms in the expression above with 
both $\lambda_{\chi,\psi}$ set to unity. From this construction it is easily seen that the $Z_1'$ interactions actually only depend upon on 3 
effective parameter combinations. For example, defining the auxiliary parameters $\epsilon_1=-s\lambda_\chi+c\lambda_\psi \delta_3$, 
$\epsilon_2=c\lambda_\psi$ and $\epsilon_3=-s\lambda_\chi \delta_1+c\lambda_\psi \delta_2$, we can re-write the general $Z_1'$ interaction above 
more simply as just 
\begin{equation}
{g\over {c_w}} \sqrt{{5x_w}\over {3}} \Bigg[\epsilon_1 {Q_\chi\over {2\sqrt {10}}}+\epsilon_2 {Q_\psi\over {2\sqrt {6}}}+\epsilon_3 
\sqrt{3\over {5}} {Y\over {2}} \Bigg]\,.
\end{equation}
This same form holds for both the triple GKM scenario considered here as well as in the more limited case of GKM mixing within the previously 
studied $U(1)_\theta$ model and also in the case when GKM is completely absent. The only difference in these three cases is then seen to be 
in the allowed ranges of the quantities $(\epsilon_i)$ which are found to be somewhat different in the triply GKM mixed case considered here. 
This demonstrates that only 3 parameters actually determine the physical $Z'_1$ couplings even though many more (calculable) parameters actually 
appear in the original Lagrangian.

\begin{figure}[htbp]
\centerline{
\includegraphics[width=9.0cm,angle=90]{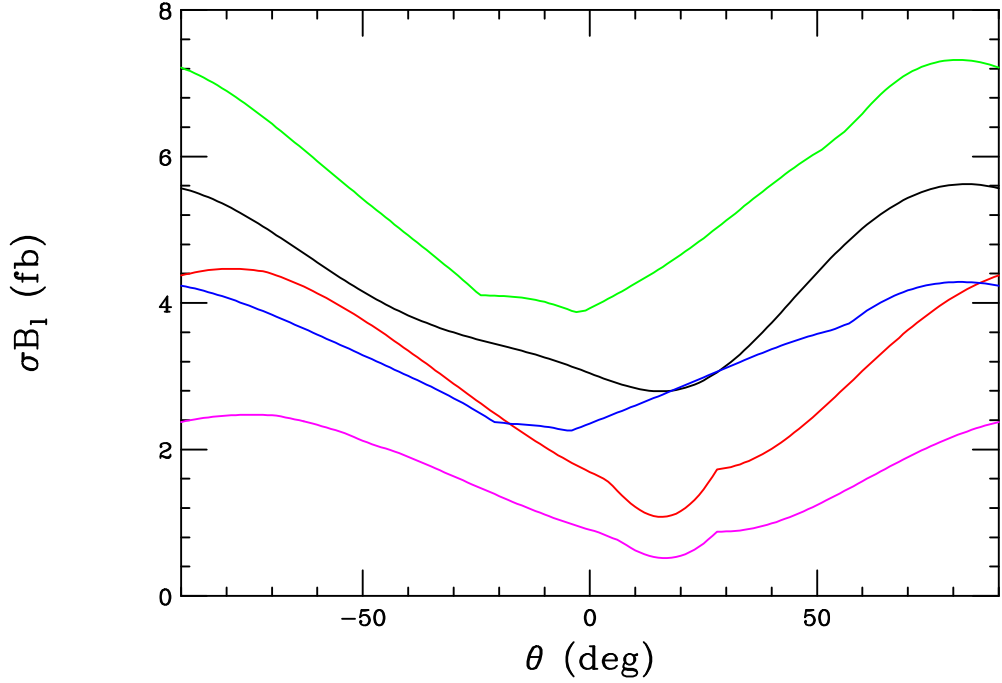}}
\vspace*{0.5cm}
\caption{Comparison of the $\sigma_{Z'} \cdot B_\ell$ values as a function of $\theta$ at the 7 TeV LHC with $M_{Z'}$=1.5 TeV for the five scenarios 
discussed in the text. The solid curve corresponds to the standard prediction without any GKM included. The red(green) curve corresponds to the 
minimum(maximum) value obtained when GKM is present for any of the 68 models. Kinks in the curves correspond to changes in which model produces 
the extreme value. The corresponding magenta(blue) curves are the maximum(minimum) values for the case when GKM occurs and additional singlets 
are also present.}
\label{fig4}
\end{figure}

To proceed further with our numerical exercise, it is sufficient to examine a specific example. First consider 
a `standard' $E_6$ $Z'$, in the absence of any GKM with a mass of 1.5 TeV 
being produced at the $\sqrt s=7$ TeV LHC{\footnote {Assuming somewhat different $Z'$ masses or going to the 8 or 14 TeV LHC will not alter 
the qualitative nature of the results obtained below.}}. Varying {\it only} the single parameter $\theta$ one can ask for that value for 
which $\sigma_{Z'}\cdot B_\ell$ is minimized or maximized. (Note that in performing these calculations we will employ the CTEQ6.6M 
pdfs{\cite{Pumplin:2002vw}} and will include an approximate  NNLO K-factor{\cite{Melnikov:2006kv}} as well as both QCD and QED corrections to 
the $Z'$ partial decay widths.) A quick scan shows that the minimum (maximum) value, $\sigma_{Z'}\cdot B_\ell \simeq 2.79(5.62)$ fb, occurs 
when $\theta \simeq 15.54(82.76)^o$. Now if we allow for triple GKM within any of the 68 models above, what is the new minimum (maximum) value of 
$\sigma_{Z'}\cdot B_\ell$?  In such a situation, as $\theta$ is varied different members of the set of 68 models may lead to this minimum or 
maximum value. In this case we find these values 
to now be $\simeq 1.08(7.32)$ fb occurring when $\theta \simeq 15.80(81.00)^o$. If we further allow for 
the presence of up to two sets of additional singlet fields, repeating this analysis we instead find this minimum (maximum) to be 
$\simeq 0.51(4.29)$ fb and which now occurs at $\theta \simeq 16.55(81.57)^o$. Note that the effect of addition singlets is to {\it suppress} the 
maximum production rate; this is to be expected as the enhanced RGE running reduces the sizes of the gauge couplings. These results can be seen 
even more clearly by examining Fig.~\ref{fig4} which shows the $\theta$ dependence of the value of $\sigma_{Z'}\cdot B_\ell$ for these five cases 
as just discussed. The kinks we observe in this plot are due to transitions between the models which lead to the different $\sigma_{Z'}\cdot B_\ell$ 
extremum. As we scan over the values of $\theta$ the difference of the two extrema are never found to be more than an order of magnitude. 
These $\sigma_{Z'}\cdot B_\ell$ values are not seen to be vastly different from those found in Ref.{\cite {GKM}} from which we 
conclude (unfortunately?) that going to the triple GKM scenario considered here from the previously considered $Z'_\theta$ model does not lead to 
any drastic modifications in the rate of dileptons arising from $Z'$ production at the LHC.

\section{Summary and Conclusions}

The $E_6$SSM scenario can address many of the outstanding issues associated with the MSSM and predicts numerous new states at the $\sim$ TeV mass 
scale that may be observed at the LHC, among them being new gauge bosons. In this paper we have examined the effect of gauge kinetic mixing between 
the three $U(1)$ fields that are generally present in the low energy sector of this model: the usual SM $U(1)_Y$ and the $U(1)_{\chi,\psi}$ 
associated with these new $Z'$s. To this end we examined a set of 68 toy models (and their extensions which included SM singlets) and analyzed the 
resulting possible affect of GKM on the $Z'$ couplings in each case as a function of the $Z'_{\chi,\psi}$ mixing angle, $\theta$.  As an example of 
the potential influence of GKM on $Z'$ production at the 7 TeV LHC in these models, for each value of this mixing angle the set of toy models were 
scanned to determine both the largest and smallest possible event rate for a fixed 1.5 TeV $Z'$ mass. Within the set of models examined we found 
that GKM did not lead to any drastic alterations in the dilepton production rate with the difference between the maximum and minimum values we 
obtained always being less than one order of magnitude. 

Hopefully the new physics associated with the $E_6$SSM scenario may soon be discovered at the LHC.

\newpage

\noindent{\Large\bf Acknowledgments}\\

The author would like to thank J.L. Hewett for discussions related to this work.

%
%%%%%%%%%%%%%%%%%%--- References
%%%%%%%%%%%%%%%%%%%%%%%%%%%%%%%%%%%%%%%%%%%%%%%%%%%%%%%
\def\IJMP #1 #2 #3 {Int. J. Mod. Phys. A {\bf#1},\ #2 (#3)}
\def\MPL #1 #2 #3 {Mod. Phys. Lett. A {\bf#1},\ #2 (#3)}
\def\NPB #1 #2 #3 {Nucl. Phys. {\bf#1},\ #2 (#3)}
\def\PLBold #1 #2 #3 {Phys. Lett. {\bf#1},\ #2 (#3)}
\def\PLB #1 #2 #3 {Phys. Lett. B {\bf#1},\ #2 (#3)}
\def\PR #1 #2 #3 {Phys. Rep. {\bf#1},\ #2 (#3)}
\def\PRD #1 #2 #3 {Phys. Rev. D {\bf#1},\ #2 (#3)}
\def\PRL #1 #2 #3 {Phys. Rev. Lett. {\bf#1},\ #2 (#3)}
\def\PTT #1 #2 #3 {Prog. Theor. Phys. {\bf#1},\ #2 (#3)}
\def\RMP #1 #2 #3 {Rev. Mod. Phys. {\bf#1},\ #2 (#3)}
\def\ZPC #1 #2 #3 {Z. Phys. C {\bf#1},\ #2 (#3)}

\end{document}